\documentclass[a4paper,groupedaddress,onecolumn,final,nofootinbib,nobibnotes,floats,oneside,hyperref]{revtex4}%
\usepackage{graphicx}

\begin{document}

\title{Effect of platy- and leptokurtic distributions in the
random-field Ising model: Mean field approach}
\author{S\'{\i}lvio M. Duarte Queir\'os$^{1}$}
\thanks{Corresponding author: sdqueiro@gmail.com}
\author{Nuno Crokidakis$^{2}$}
\thanks{nuno@if.uff.br}
\author{Diogo O. Soares-Pinto$^{3}$}
\thanks{dosp@cbpf.br}

\address{
$^{1}$Unilever R\&D Port Sunlight \\
Quarry Road East Wirral, CH63 3JW UK \\
$^{2}$Instituto de F\'{\i}sica - Universidade Federal Fluminense \\
Av. Litor\^anea s/n \\
24210-340 \hspace{5mm} Niter\'oi - RJ \hspace{5mm} Brazil \\
$^{3}$Centro Brasileiro de Pesquisas F\'{\i}sicas \\
Rua Dr Xavier Sigaud 150 \\
22290-180 \hspace{5mm} Rio de Janeiro - RJ \hspace{5mm} Brazil}

\date{30th June 2009}

\begin{abstract}
\noindent The influence of the tail features of the local magnetic field
probability density function (PDF) on the ferromagnetic Ising model is
studied in the limit of infinite range interactions. Specifically, we assign
a quenched random field whose value is in accordance with a generic
distribution that bears platykurtic and leptokurtic distributions depending
on a single parameter $\tau < 3$ to each site. For $\tau< 5/3$, such
distributions, which are basically Student-$t$ and $r$-distribution extended
for all plausible real degrees of freedom, present a finite standard
deviation, if not the distribution has got the same asymptotic power-law
behavior as a $\alpha $-stable L\'{e}vy distribution with $\alpha = (3 -
\tau )/(\tau - 1)$. For every value of $\tau $, at specific temperature and
width of the distribution, the system undergoes a continuous phase
transition. Strikingly, we impart the emergence of an inflexion point in the
temperature-PDF width phase diagrams for distributions broader than the
Cauchy-Lorentz ($\tau = 2$) which is accompanied with a divergent free
energy per spin (at zero temperature).
\end{abstract}

\maketitle

\section{Introduction}

Disorder is ubiquitous in Nature. Regarding materials and their statistical
properties, disordered magnetic systems have been systematically studied in
condensed matter and statistical physics. From a theoretical point of view,
the most studied case has certainly been the Random Field Ising Model (RFIM)~%
\cite{binder_review,dotsenkobook}, because of its simplicity as a frustrated
system and relevancy to experiments~\cite{belangerreview, birgeneau} which
has been quite boosted after the identification of the RFIM with diluted
antiferromagnets in the presence of a uniform magnetic field \cite%
{belangerreview,fishmanaharony,pozenwong,cardy} and several ferromagnetic
compounds as well~\cite{belangerreview,birgeneau,kushauerkleemann}.

In order to generate the local random field, both the Gaussian and the
bimodal probability density function (PDF) have intensively been used \cite%
{machta00, middleton,hartmann}. Nevertheless, controversy over the order of
the low-temperature phase transition has still been at the helm of several
discussions. On one hand, a high temperature series expansion up to $15$th
order showed a continuous phase transition for both the Gaussian and the
bimodal PDF~\cite{gofman}. On the other hand, from an exact determination of
the ground states in higher dimensions ($d=4$), Swift \textit{et al}~\cite%
{swift} found a discontinuous phase transition for the bimodal random field,
whereas for $d=3$ dimensions and the Gaussian distribution the transition is
continuous. By applying the Wang-Landau algorithm \cite{wang_landau}, recent
simulations on 3D lattices claimed the discovery of first-order-like
features in the strongly disordered regime for both those PDFs~\cite%
{hernandez,machta06}.

As an alternative to the above mentioned approaches, there is the mean field
theory which can present a good qualitative agreement with some short-range
interaction models and experiments. Once more, the Gaussian and the bimodal
PDF have been widely investigated \cite{schneiderpytte, aharony} as well as
related distributions such as the trimodal~\cite{mattis,kaufman} and the
double-Gaussian~\cite{nuno08} or the treble-Gaussian~\cite{nuno09}. In the
Gaussian RFIM case, the phase diagram only presents continuous phase
transitions~\cite{schneiderpytte}, whereas in the bimodal case the phase
diagram presents a continuous phase transition for high temperatures and low
random-field intensities and for low temperatures and high random-field
intensities a first-order transition arises therefrom~\cite{aharony}. In
other more elaborated cases a rich critical behavior can be found for finite
temperatures as it has been recently conveyed in \cite{nuno08,nuno09}.
Accordingly, we can understand that the choice of the local random field PDF
is of crucial importance for a good theoretical description of real systems.
In this particular context and based on the identification of the RFIM with
diluted antiferromagnets in a uniform field, for which the local random
fields are expressed in terms of quantities that vary in both signal and
magnitude \cite{fishmanaharony,cardy}, the use of continuous PDFs has
demonstrated to be a very promising approach~\cite{nuno08,nuno09}.

The utilization of Dirac Delta and Gaussian related distributions is much
supported on the easiness of the analytical treatment of the subsequent
equations as well as the pervasiveness of the Gaussian distribution.
Although the Gaussian was assumed for many generations as the ``natural
distribution'', in the last decades the concept of (asymptotic)
scale-invariance of probability density functions have abundantly emerged~%
\cite{viscek}. In the realm of disordered systems, PDFs different to the $n$%
-Gaussian or the $n$-Dirac Delta were used to explain the critical behavior
of several compounds. For instance, PDFs with very fat tails were introduced
to analyze organic charge-transfer compounds like: N-methyl-phenazium
tetra-cyanoquinodimethanide (NMP-TCNQ), quinolinium-(TCNQ)$_{2}$,
acridinium-(TCNQ)$_{2}$ and phenazine-TCNQ, as first reported in Refs.~\cite%
{fat-materials}. Conversely, a sub-Gaussian distribution was used to account
for the magnetic properties and the critical behavior of poly(metal
phosphinates)~\cite{slim-materials}. Last but not least, as was proven by
Gosset~\cite{student}, asymptotic scale invariant distributions can be
derived from the Gaussian distribution when finite elements are taken into
account so that finite and scale-dependent systems can be treated as
infinite and (asymptotically) scale-independent. Therefore, the study of
more general continuous PDFs turns up very interesting as it furnishes a
more widespread picture of disordered magnetic systems than the
distributions used up to now. With such a goal in mind, we study herein the
aftermath of applying a more general family of continuous PDFs in the mean
field RFIM. Explicitly, our PDF reproduces the $r$- and $t$-distributions
for real degrees of freedom. For specific values of the triplet composed of
the degree of freedom, the temperature and the PDF width, our results show
that the system experiences a continuous phase transition that does not
dependent on the finiteness of the standard deviation and the scale behavior
(dependence or independence) of the random field. Moreover, for PDFs fatter
than the Cauchy-Lorentz, we determine the emergence of an inflexion point in
the temperature versus PDF width phase diagrams that coexists with a
divergence at zero temperature of the free energy per spin.

%%%%%%%%%%%%%%%%%%%%%%%%%%%%%%%%%%%%%%%%%%%%%%%%%%%%%%%%%%%%%%%%%

\section{The Model}

The infinite-range-interaction Ising model in the presence of an external
random magnetic field is defined in terms of the Hamiltonian,

\begin{equation}
\mathcal{H}=-\frac{J}{N}\sum_{(i,j)}S_{i}S_{j}-\sum_{i}H_{i}S_{i}~,
\label{1}
\end{equation}%
\noindent where the sum $\sum_{(i,j)}$ runs over all distinct pairs of spins
$S_{i}=\pm 1$ ($i=1,2,...,N$). The random fields $\{H_{i}\}$ are quenched
variables and ruled by a PDF that is defined by a parameter $\tau $ (generic
degree of freedom). For $\tau <1$,
\begin{equation}
P_{i}(H_{i})=\sqrt{\frac{1-\tau }{\pi }\mathcal{B}_{\tau }}\frac{\;\Gamma %
\left[ \frac{5-3\tau }{2\left( 1-\tau \right) }\right] }{\;\Gamma \left[
\frac{2-\tau }{1-\tau }\right] }[1-\mathcal{B}_{\tau }(1-\tau )H_{i}^{2}]^{%
\frac{1}{1-\tau }},  \label{2}
\end{equation}%
(\noindent with $\left\vert H\right\vert \leq \left[ \mathcal{B}_{\tau
}(1-\tau )\right] ^{-1/2}$) which is the generalized $r$-distribution, and
for $\tau >1$, we have
\begin{equation}
P_{s}(H_{i})=\sqrt{\frac{\tau -1}{\pi }\mathcal{B}_{\tau }}\frac{\;\Gamma %
\left[ \frac{1}{\tau -1}\right] }{\;\Gamma \left[ \frac{3-\tau }{2\left(
\tau -1\right) }\right] }[1-\mathcal{B}_{\tau }(1-\tau )H_{i}^{2}]^{\frac{1}{%
1-\tau }},  \label{3}
\end{equation}%
\noindent which is the generalized Student-$t$ distribution. By \textit{%
generalized} we mean that the degrees of freedom, $m$ and $n$, of $t$- and $%
r $-distributions are extended to the entire domain of feasible real values
according to the relations $\tau =\left( m+3\right) /\left( m+1\right) $ $%
\left[ m\geq 0\right] $ and $\tau =\left( n-4\right) /\left( n-2\right) $ $%
\left[ n\geq 2\right] $, respectively. In Eqs.~(\ref{2}) and (\ref{3}), $%
\Gamma \lbrack .]$ is the Gamma function and $\mathcal{B}_{\tau }$ is given
by
\begin{equation}
\mathcal{B}_{\tau }=\frac{1}{(3-\tau )\,\omega ^{2}},  \label{4}
\end{equation}%
where $\omega $ is the width of the PDF. For $\tau <5/3$ the width and the
standard deviation, $\sigma $, are related by
\begin{equation}
(5-3\tau )\,\sigma ^{2}=(3-\tau )\,\omega ^{2}~.  \label{5}
\end{equation}%
Alternatively, the functional form of Eqs.~(\ref{2}) and (\ref{3}) can be
obtained by optimizing the entropic form presented in \cite{tsallis} by
applying the concept of escort distribution, $p\left( H\right) \equiv
P^{\tau }\left( H\right) /\int P^{\tau }\left( H\right) \,dH$ \cite%
{souza_ct,beckbook}, and for that is many times called $q$\emph{-Gaussian}.
In this case, $\omega ^{2}$ plays the role of the constraint, $\int H^{2}p\left(
H\right) \,dH=\omega ^{2}$, which is always finite for $\tau <3$ with the
corresponding Lagrange multiplier given by Eq. (\ref{4}). Expressly, $\omega
^{2}$ represents the standard deviation of the escort distribution and it is
finite even when the distribution \textit{per se} has got a divergent
standard deviation, $\int H^{2}P\left( H\right) \,dH=\sigma ^{2}$.
Therefore, it represents a way of appraizing the broadness of the
distribution and this is the reason why we named $\omega $ width. Recently, $%
P_{i\left( s\right) }\left( H\right) $ has also been coined \emph{%
generalized Lorentzian} \cite{hilhorst}. Although we acknowledge both
nomenclatures we use the traditional terminology of $r-$ and $t-$
distributions that is quite well established in the Statistics community
since a long time. The PDF defined in Eqs.~(\ref{2}) and (\ref{3}) is symmetrical around $H=0$ and represents a family of continuous distributions that recovers some well-known distributions using appropriate limits, namely:

\begin{itemize}
\item the uniform distribution, for $\tau \rightarrow -\infty $;

\item compact support distributions (limited), for $\tau <1$;

\item the Gaussian distribution, for $\tau \rightarrow 1$;

\item the Cauchy-Lorentz distribution, for $\tau =2$.

\item Dirac Delta, for every $\tau <3$ and $\omega\rightarrow 0$.
\end{itemize}

%%%%%%%%%%%%%%%%%%%%%%%%%%%%%%%%%%%%%%%%%%%%%%%%%%%%%%%%%%%%%%%%%%%%%%%%
\begin{figure}[t]
\begin{center}
\vspace{1.0cm} \includegraphics[width=0.47\textwidth,angle=0]{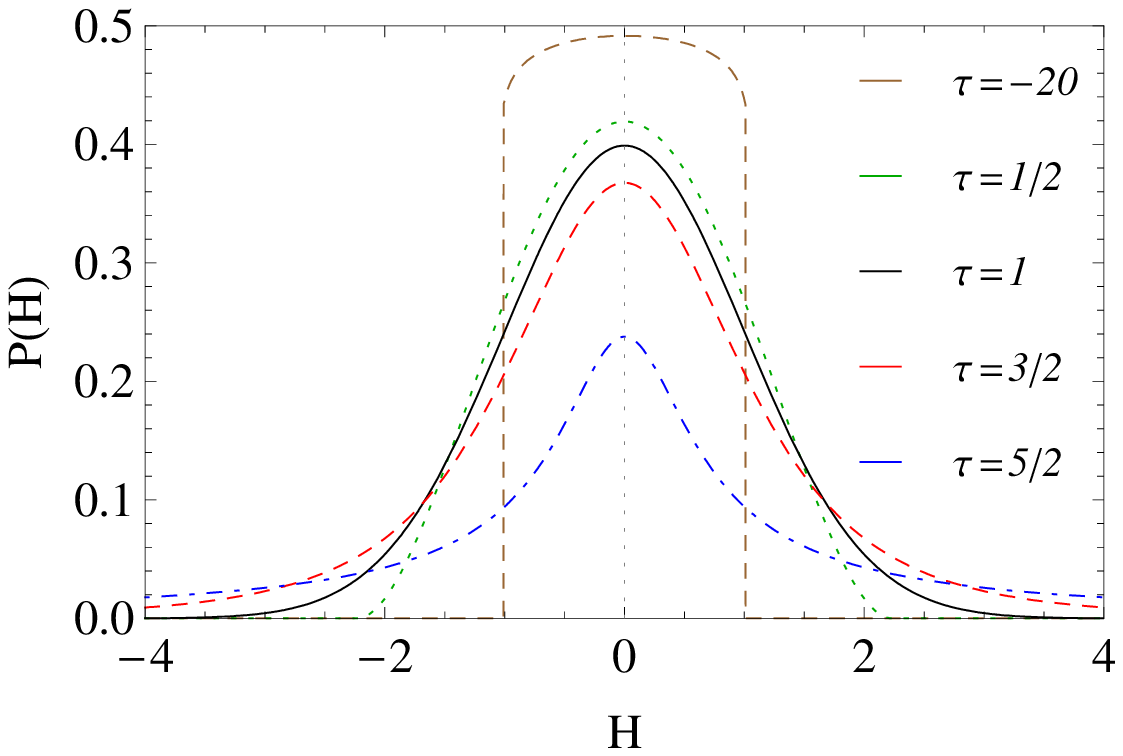}
\hspace{0.5cm} \includegraphics[width=0.45%
\textwidth,angle=0]{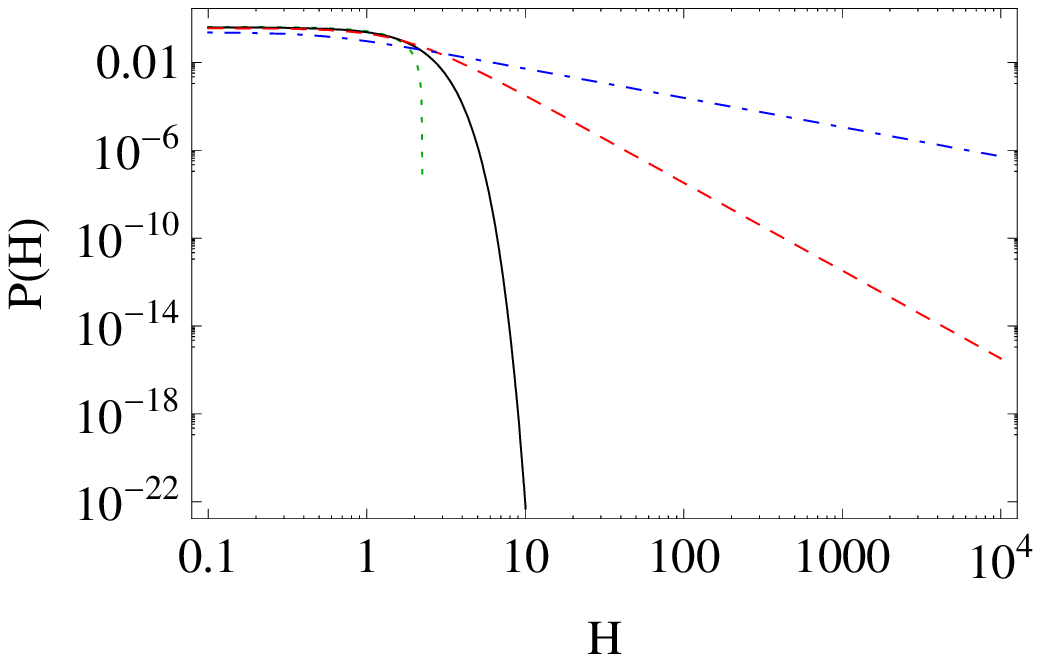}
\end{center}
\caption{(Color online) Random-field probability distributions for some
values of the parameter $\protect\tau $ (from botton to top: $\protect\tau %
=5/2,3/2,1,1/2$ and $-20$), in the normal (left panel) and log-log scale
(right panel). We have used $\protect\omega =1$ in all cases.}
\label{Fig1}
\end{figure}
%%%%%%%%%%%%%%%%%%%%%%%%%%%%%%%%%%%%%%%%%%%%%%%%%%%%%%%%%%%%%%%%%%%%%%%%%

To boot, the functional form (\ref{3}) is an asymptotic power-law decaying
PDF with finite standard deviation for $1<\tau <5/3$ and an asymptotic
power-law decaying PDF, but with infinite standard deviation instead. In
both cases the decay exponent is equal to $2/(\tau -1)$. The latter case is
also capable of reproducing the tail behavior of $\alpha $-stable L\'{e}vy
distributions%
\[
\mathcal{L}_{\alpha }\left( H\right) =\int_{-\infty }^{\infty }\exp \left[
-a\,\left\vert k\right\vert ^{\alpha }+i\,k\,H\right] \,dk,
\]
with $\alpha =\left( 3-\tau \right) /\left( \tau -1\right) $ and broadness $%
a $, whose escort-distribution has got a finite width as well. For the case
of the Cauchy-Lorentz, $\alpha =1$ ($\tau =2$) [the only case for which L%
\'{e}vy distributions are explicitly defined in real space], the parameter $%
a $ is equal to width $\omega $. Accordingly, if we bear in mind the
previous work by Aharony \cite{aharony},\ we can hold that our enquiry also
sheds light on the low temperature behavior of the random-field Ising model
with the local magnetic field associated with a $\alpha $-stable L\'{e}vy
distribution. In Fig.~\ref{Fig1}, we depict PDFs (\ref{2}) and (\ref{3}) for
some values of $\tau $. Regarding the \textit{kurtosis},%
\begin{equation}
\kappa \equiv \frac{\left\langle H^{4}\right\rangle }{\left\langle
H^{2}\right\rangle ^{2}},
\end{equation}%
the distribution is \textit{platykurtic}, $\kappa <3$, for $\tau <1$ or
\textit{leptokurtic}, $\kappa >3$, for $\tau >1$. At this point it is
important to stress that, as it has been made until now, in spite of being
able to present non-mesokurtic distributions the combination of Gaussians
results in asymptotic scale-dependent distributions.

From the free energy, $F(\{H_{i}\})$, associated with a given realization of
site fields, $\{H_{i}\}$, we calculate the quenched average, $%
[F(\{H_{i}\})]_{H}$,
\begin{equation}
\lbrack F(\{H_{i}\})]_{H}=\int \prod_{i}[dH_{i}P(H_{i})]F(\{H_{i}\})~.
\label{6}
\end{equation}%
The general mean field result of the free energy per spin, in terms of any
PDF of the random fields, is well-known \cite{schneiderpytte,aharony}, and
is given by
\begin{equation}
f=\frac{J^{2}}{2}m^{2}-\frac{1}{\beta }\langle\,\log [2\cosh \beta
(Jm+H)]\rangle_{H}  \label{7}
\end{equation}%
\noindent and the magnetization is given by,
\begin{equation}
m=\langle\,\mathrm{tanh}[\beta (Jm+H)]\rangle_{H},  \label{8}
\end{equation}%
\noindent where $\langle\,\ldots\rangle_{H}$ stands for averages over
realizations of the disorder, \textit{i.e.},
\[
\langle\,...\rangle_{H}=\int_{-\infty }^{+\infty }dH\;P(H)\;(...)~.
\]

Close to a continuous transition between ordered and disordered phases, the
magnetization $m$ is small. So, we can expand Eq.~(\ref{8}) in powers of $m$%
,
\begin{equation}
m=Am+Bm^{3}+Cm^{5}+\mathcal{O}(m^{7}),  \label{9}
\end{equation}%
\noindent where the coefficients are given by
\begin{eqnarray}  \label{13}
A &=&\beta J\{1-\rho _{1}\},  \label{11} \\
B &=&-\frac{(\beta J)^{3}}{3}\{1-4\rho _{1}+3\rho _{2}\},  \label{12} \\
C &=&\frac{(\beta J)^{5}}{15}\{2-17\rho _{1}+30\rho _{2}-15\rho _{3}\},
\end{eqnarray}%
\noindent with
\[
\rho _{k}=\langle \,\tanh ^{2k}(\beta H)\rangle _{H}~.
\]%
\noindent With the aim of finding the continuous critical frontier we set $%
A=1$, provided that $B<0$. If a first-order critical frontier also occurs,
the continuous line must end when $B=0$; in such cases, the continuous and
the first-order critical frontiers converge at a tricritical point, whose
coordinates are obtained by solving the equations $A=1$ and $B=0$, on
condition that $C<0$. Thus, for $A=1$, we obtain
\begin{equation}
\frac{kT}{J}=1-\langle \,\tanh ^{2}(\beta H)\rangle _{H}~.  \label{15}
\end{equation}

In the following section, we discuss the role of PDFs (\ref{2}) and (\ref{3}%
) when they are considered in the formulae presented in this section. Our
survey includes the analysis of the phase diagrams for the whole domain of $%
\tau $.

%%%%%%%%%%%%%%%%%%%%%%%%%%%%%%%%%%%%%%%%%%%%%%%%%%%%%%%%%%%%%%%%%%%%%%%%%

\section{Finite Temperature Analysis}

Following the above presented results, we proceed by calculating the
critical frontiers of the model when the temperature is different from zero.
In the RFIM, we have a single transition between the two possible phases of
the magnetization: the ferromagnetic phase ($m\neq 0$) and the paramagnetic
phase ($m=0$). The critical frontier separating these two phases is found by
solving Eq.~(\ref{15}). On account of the fact that Eq.~(\ref{15}) is
analytically unsolvable, we have been compelled to solve it by numerical
means using the Global Adaptative Strategy algorithm \cite{kromer} that has
been proven as the best (\textit{i.e.}, fast and accurate) numerical
integration procedure for smooth integrands \cite{malcom}.

%%%%%%%%%%%%%%%%%%%%%%%%%%%%%%%%%%%%%%%%%%%%%%%%%

\subsection{Platykurtic case: $\protect\tau <1$}

Let us denote $f_{i}$ and $m_{i}$ as the free energy and the magnetization
for this regime of $\tau $, respectively. Thus, Eqs.~(\ref{7}) and (\ref{8})
become,
\begin{equation}
f_{i}=\frac{J^{2}}{2}m_{i}^{2}-\frac{1}{\beta }\int_{-\frac{1}{\sqrt{%
\mathcal{B}_{\tau }(1-\tau )}}}^{\frac{1}{\sqrt{\mathcal{B}_{\tau }(1-\tau )}%
}}dH\,\,P_{i}(H)\log [2\cosh \beta (Jm_{i}+H)],  \label{17a}
\end{equation}%
and%
\begin{equation}
m_{i}=\int_{-\frac{1}{\sqrt{\mathcal{B}_{\tau }(1-\tau )}}}^{\frac{1}{\sqrt{%
\mathcal{B}_{\tau }(1-\tau )}}}P_{i}(H)\tanh \beta (Jm_{i}+H)~,  \label{17b}
\end{equation}%
\noindent where $P_{i}(H)$ is given by the PDF in Eq.~(\ref{2}). The
continuous critical frontier has been found when we have solved Eq.~(\ref{15}%
). For all solutions obtained, we have calculated a negative value of $B$,
Eq.~(\ref{12}), which has confirmed the continuous character of the phase
transition.

%%%%%%%%%%%%%%%%%%%%%%%%%%%%%%%%%%%%%%%%%%%%%%%%%%%%%%%%%%%%%%%%%%%%%%%%
\begin{figure}[t]
\begin{center}
\vspace{1.0cm} \includegraphics[width=0.6\textwidth,angle=0]{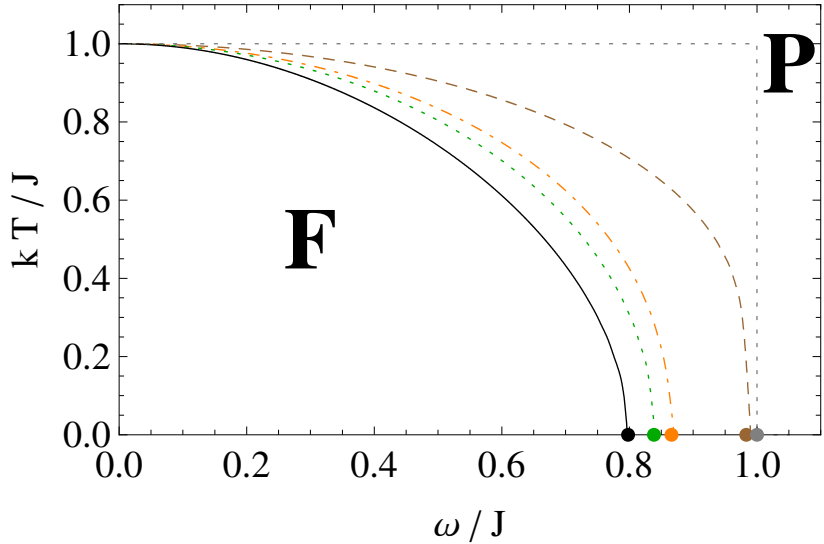}
\end{center}
\caption{(Color online) Phase diagram of the model, in the plane temperature
vs $\protect\omega $ (in units of $J$), for some values of the parameter $%
\protect\tau <1$. The grey dotted line is for $\protect\tau =-\infty $ \ and
$\protect\omega \left( T=0\right) =J$; the brown dashed line is for $\protect%
\tau =-20$ and $\protect\omega \left( T=0\right) =0.9831\ldots J$; the
dot-dashed orange line is for $\protect\tau =0 $ and $\protect\omega \left( T=0\right)=0.8660\ldots J$; the dotted green line is for $\protect\tau =1/2$ and $%
\protect\omega \left( T=0\right) =3\protect\sqrt{5/64}J$ and the black full
line is the Gaussian case with $\protect\omega \left( T=0\right) =\protect%
\sqrt{2/\protect\pi }J$. We can observe continuous phase transitions
between the Ferromagnetic (\textbf{F}) and the Paramagnetic (\textbf{P})
phases for all values of $\protect\tau $. The points onto the $\protect%
\omega /J$ axis were exactly calculated through a zero-temperature analysis
[section IV.A] where from we can see a good agreement between the analytical
and numerical results which by interpolation indicates discrepancies never
greater than 1\%.}
\label{Fig2}
\end{figure}
%%%%%%%%%%%%%%%%%%%%%%%%%%%%%%%%%%%%%%%%%%%%%%%%%%%%%%%%%%%%%%%%%%%%%%%%%

If a first-order transition existed as well, the critical frontier would be
found by equalizing the free energy at each side of this line, i.e., $%
f(m=0)=f(m\neq 0)$. Using this procedure, we have numerically determined the
critical frontiers separating the paramagnetic and ferromagnetic phases, for
typical values of $\tau <1$. We have confirmed that the above coefficient $B$%
, Eq.~(\ref{12}), is always negative. The phase diagram is shown in Fig.~\ref%
{Fig2}, on the plane defined by the temperature, $T$, and the PDF width, $%
\omega $ (both in units of $J$), for some typical values of $\tau <1$. In
that figure, the lines represent the numerical solution of Eq.~(\ref{15}),
whereas the points were analytically obtained through a zero-temperature
analysis, which is going to be discussed in the next section. Notice that
the ferromagnetic phase is reduced by increasing the parameter $\tau$ from $%
\tau=-\infty$ to $\tau=1$ as shown in Fig.~\ref{Fig2}, and for the maximum
value for $r$-distributions, $\tau =1$, we recover the simple phase diagram
of the Gaussian distribution \cite{schneiderpytte}.

\subsection{Leptokurtic case: $\protect\tau >1$}

Analogously to the platykurtic case, we denote $f_{s}$ and $m_{s}$ as the
free energy and the magnetization per spin for this regime of $\tau $. The
expansion of the magnetization Eq.~(\ref{9}) is valid for this case as well,
but the averages over the disorder $\langle\,\ldots \rangle_{H}$ must be
made according to PDF (\ref{3}),
\begin{eqnarray}
f_{s} & = & \frac{J^{2}}{2}m_{s}^{2}-\frac{1}{\beta }\int_{-\infty}^{+%
\infty}dH\,\,P_{s}(H)\log [2\cosh \beta (Jm_{s}+H)], \\
m_{s} & = &\int_{-\infty}^{+\infty}P_{s}(H)\tanh \beta (Jm_{s}+H)~,
\end{eqnarray}%
\noindent where in this case the integration limits are taken in the range $%
(-\infty,+\infty)$.

%%%%%%%%%%%%%%%%%%%%%%%%%%%%%%%%%%%%%%%%%%%%%%%%%%%%%%%%%%%%%%%%%%%%%%%
\begin{figure}[t]
\begin{center}
\vspace{1.0cm} \includegraphics[width=0.6\textwidth,angle=0]{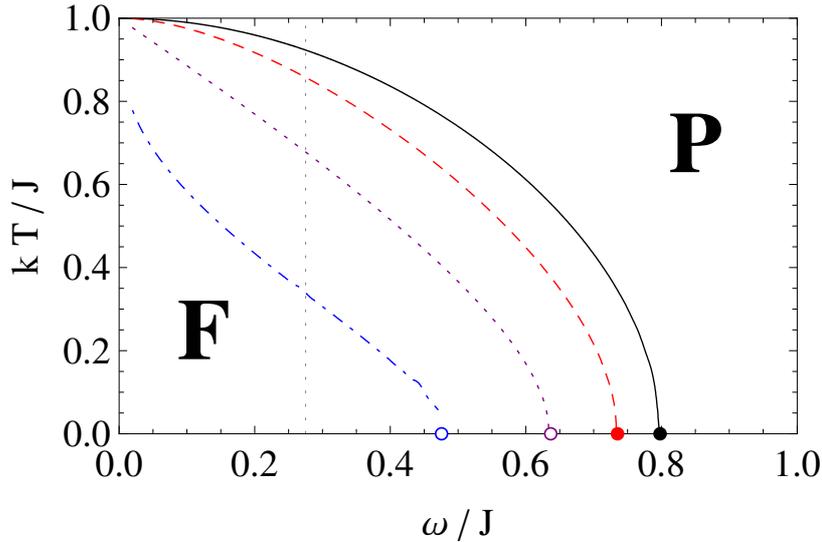}
\end{center}
\caption{(Color online) Phase diagram of the model, in the plane temperature
versus $\protect\omega$ (in units of $J$), for some values of the parameter $%
\protect\tau >1$. The black full line is the Gaussian case with $\protect%
\omega \left( T=0\right) =\protect\sqrt{\frac{2}{\protect\pi }}J$; red
dashed line is for $\protect\tau =3/2$ \ and $\protect\omega \left(
T=0\right) =\frac{4}{\protect\sqrt{3}\protect\pi }J$; the purple dotted line
is for $\protect\tau =2$ and $\protect\omega \left( T=0\right) =\frac{2}{%
\protect\pi }J$; the dot-dashed blue line is for $\protect\tau =5/2$ with $%
\protect\omega \left( T=0\right) =0.4754\ldots J$. We can observe continuous
phase transitions between the Ferromagnetic (\textbf{F}) and the
Paramagnetic (\textbf{P}) phases for all values of $\protect\tau $. The
points represent the results obtained by the zero-temperature analysis.
Notice the change in the concavity of the critical frontier for large values
of $\protect\tau $ ($>2.0 $). The vertical dashed line is $\protect\omega %
=0.275 J$ which is close to the inflexion point of the critical line for $\protect%
\tau =5/2$. In this figure, we have distinguished the points with finite
free energy per spin from the points with a divergent free energy per spin
representing the latter by empty circles. Again, we can observe a good
agreement between the numerics and the expansion at $T=0$. The difference
between the analytical approximation and interpolation is again never greater than 1\%.}
\label{Fig3}
\end{figure}
%%%%%%%%%%%%%%%%%%%%%%%%%%%%%%%%%%%%%%%%%%%%%%%%%%%%%%%%%%%%%%%%%%%%%%%%%

By considering PDF~(\ref{3}), the above presented procedure for the
determination of the critical frontiers can be employed once more.\textit{\ }%
In other words, Eq.~(\ref{15}) provide the continuous critical line of the
phase diagram. Using this procedure, we have numerically evaluated the
critical frontiers separating the paramagnetic and ferromagnetic phases for
typical values of $\tau >1$. Like the platykurtic case, the leptokurtic case
has only given negative values of $B$, $i.e.$, no other than continuous
phase transition occurs. The phase diagram is shown in Fig. \ref{Fig3}, on
the plane formed by the temperature and the PDF width $\omega $ (in units of
$J$), for some specific values of $\tau >1$. Still, the lines represent
numerical solutions of Eq.~(\ref{15}), while at the same time the points
were analytically obtained through a zero-temperature analysis, which is
going to be discussed shortly. As we have perceived in the platykurtic case,
the ferromagnetic phase is reduced by augmenting $\tau $. Similar behavior
was found in the Gaussian \cite{schneiderpytte} and the double-Gaussian RFIM
\cite{nuno08} by increasing the standard deviation of such PDFs. However, a
chief difference emerges. For distributions with fatter tails than the
Cauchy-Lorentz PDF, the concavity of the critical frontier changes in the
high-temperature region. So far as we are aware, this is the first time that
such a change is observed in the mean-field RFIM phase diagram.

%%%%%%%%%%%%%%%%%%%%%%%%%%%%%%%%%%%%%%%%%%%%%%%%%%%%%%%%%%%%%%%%%%%%%%%%%%

\section{Zero Temperature Analysis}

Moving forwards, we now consider the phase diagram of the model at zero
temperature. As in the finite-temperature case, we evolve twofold: the
platykurtic case and the leptokurtic case, $\tau <1$ and $\tau >1$,
respectively.

\subsection{Platykurtic case: $\protect\tau <1$}

In the limit $T \rightarrow 0$, the free energy and magnetization become$%
^{1} $\footnotetext[1]{%
For the purpose of obtaining the following expressions we made use of the
integrals presented in Ref.~\cite{gradshteyn}.}, respectively,
\begin{eqnarray}  \label{18}
f_{i} &=&\frac{\Gamma \left[ \frac{5-3\tau }{2\left( 1-\tau \right) }\right]
}{2(2-\tau )\Gamma \left[ \frac{2-\tau }{1-\tau }\right] }\sqrt{\frac{1-\tau
}{\left( 5-3\tau \right) \pi }}  \nonumber  \label{fplaty} \\
&&\left\{ 4\frac{(2-\tau )\,_{2}F_{1}\left[ \frac{1}{2},\frac{1}{\tau -1};%
\frac{3}{2};\frac{\left( 1-\tau \right) J^{2}}{(5-3\tau )\sigma ^{2}}%
m_{i}^{2}\right] J^{2}}{\sigma }m_{i}^{2}+\right.  \nonumber \\
&&(5-3\tau )\sigma ^{3}\left[ 2\left( 1-\frac{(1-\tau )J^{2}}{(5-3\tau
)\sigma ^{4}}m_{i}^{2}\right) ^{1+\frac{1}{1-\tau }}-1-\left[ \sigma
^{2}\left( \sigma ^{2}-1\right) \right] ^{^{1+\frac{1}{1-\tau }}}\right] +
\nonumber \\
&&\left. 2(2-\tau )\sqrt{\frac{5-3\,\tau }{1-\tau }}\,_{2}F_{1}\left[ \frac{1%
}{2},\frac{1}{\tau -1};\frac{3}{2};\frac{1}{\sigma ^{2}}\right]
(1-J\,m_{i})\right\} ,
\end{eqnarray}%
\noindent and%
\begin{equation}
m_{i}=2\sqrt{\frac{1-\tau }{\left( 5-3\tau \right) \pi }}\frac{\,\,\Gamma %
\left[ \frac{5-3\tau }{2\left( 1-\tau \right) }\right] }{\Gamma \left[ \frac{%
2-\tau }{1-\tau }\right] }\left( \frac{J}{\sigma }\right) \,_{2}F_{1}\left[
\frac{1}{2},\frac{1}{\tau -1};\frac{3}{2};\frac{(1-\tau )J^{2}}{(5-3\,\tau
)\sigma ^{2}}m_{i}^{2}\right] m_{i},  \label{platymag}
\end{equation}%
where $_{2}F_{1}[.,.;.;.]$ is the Gauss hypergeometric function \cite%
{hipergeometrica}. In the same way as in the finite-temperature analysis, we
expand the above magnetization (\ref{platymag}) in powers of $m_{i}$, so
that
\begin{equation}
m_{i}=a_{i}m_{i}+b_{i}m_{i}^{3}+c_{i}m_{i}^{5}+\mathcal{O}(m_{i}^{7}),
\label{20}
\end{equation}%
\noindent where%
\begin{eqnarray}
a_{i} &=&-2\left( \frac{J}{\sigma }\right) \sqrt{\frac{(1-\tau )^{3}}{%
(5-3\tau )\pi }}\frac{\Gamma \left[ \frac{5-3\tau }{2\left( 1-\tau \right) }%
\right] }{\Gamma \left[ \frac{1}{1-\tau }\right] },  \label{23} \\
b_{i} &=&-\frac{2}{3\sqrt{\pi }}\left( \frac{J}{\sigma }\right) ^{3}\left(
\frac{1-\tau }{5-3\tau }\right) ^{3/2}\frac{\Gamma \left[ \frac{5-3\tau }{%
2\left( 1-\tau \right) }\right] }{\Gamma \left[ \frac{1}{1-\tau }\right] },
\\
c_{i} &=&\frac{1}{5\sqrt{\pi }}\left( \frac{J}{\sigma }\right) ^{5}\sqrt{%
\frac{1-\tau }{(5-3\tau )^{5}}\tau }\frac{\Gamma \left[ \frac{5-3q}{2\left(
1-q\right) }\right] }{\Gamma \left[ \frac{2-\tau }{1-\tau }\right] }.
\end{eqnarray}

\noindent The continuous critical frontier at zero temperature is obtained
for $a_{i}=1 $,
\begin{equation}
\frac{\sigma }{J}=\frac{2(1-\tau )}{\sqrt{\pi }}\left( \frac{1-\tau }{%
5-3\tau }\right) ^{1/2}\frac{\Gamma \left[ \frac{5-3\tau }{2(1-\tau )}\right]
}{\Gamma \left[ \frac{1}{1-\tau }\right] },  \label{24}
\end{equation}%
\noindent providing that $b_{i}<0$, which occurs for all $\tau <1$. The
last-mentioned equation allows determining the exact point at which the
critical frontiers obtained in section 3.A reach the zero-temperature axis
(the circles in Fig. \ref{Fig2}). The zero-temperature phase diagram is
shown in Fig. \ref{Fig4}.

%%%%%%%%%%%%%%%%%%%%%%%%%%%%%%%%%%%%%%%%%%%%%%%%%%%%%%%%%%%%%%%%%%%%%%%%%

\subsection{Leptokurtic case: $\protect\tau >1$}

In this regime, PDF (\ref{3}) presents a distinct behavior for $1<\tau <5/3$
and $\tau >5/3$. Explicitly, the former case corresponds to the case in
which the standard deviation is finite and the latter to the case for which
the distribution has the same asymptotic behavior as the L\'{e}vy
distribution.

\subsubsection{Finite standard deviation: $1<\protect\tau <5/3$}

For this range of $\tau $, the free energy and the magnetization become,%
\begin{eqnarray}
f_{s} &=&\sqrt{\frac{5-3\tau }{\left( \tau -1\right) \pi }}\,\frac{\Gamma %
\left[ \frac{2-\tau }{\tau -1}\right] }{\Gamma \left[ \frac{3-\tau }{2\left(
\tau -1\right) }\right] }\,\sigma \left\{ \left[ 1-\frac{1-\tau }{5-3\tau }%
\left( \frac{J}{\sigma }\right) ^{2}m_{s}^{2}\right] ^{\frac{1}{1-\tau }%
+1}+\right.  \nonumber  \label{flepto} \\
&&\left. 2(2-\tau )\left( \frac{J}{\sigma }\right) ^{2}\,_{2}F_{1}\left[
\frac{1}{2},\frac{1}{\tau -1};\frac{3}{2};-\frac{1-\tau }{5-3\tau }\left(
\frac{J}{\sigma }\right) ^{2}m_{s}^{2}\right] m_{s}^{2}\right\} ,
\end{eqnarray}%
and
\begin{eqnarray}
m_{s} &=&2\,m_{s}\left( \frac{J}{\sigma }\right) \sqrt{\frac{\tau -1}{\left(
5-3\tau \right) \pi }}\frac{\Gamma \left[ \frac{1}{\tau -1}\right] }{\Gamma %
\left[ \frac{3-\tau }{2\left( \tau -1\right) }\right] }\;\times  \label{25}
\\
&&_{2}F_{1}\left[ \frac{1}{2},\frac{1}{\tau -1};\frac{3}{2};-\frac{1-\tau }{%
5-3\tau }\left( \frac{J}{\sigma }\right) ^{2}m_{s}^{2}\right] ~,  \nonumber
\end{eqnarray}%
respectively. Similarly to the $\tau <1$ analysis, we can expand the
magnetization $m_{s}$, Eq.~(\ref{25}), in powers of $m_{s}$,
\begin{equation}
m_{s}=a_{s}m_{s}+b_{s}m_{s}^{3}+c_{s}m_{s}^{5}+\mathcal{O}(m_{s}^{7}),
\label{26}
\end{equation}%
\noindent where,
\begin{eqnarray}
a_{s} &=&2\,\left( \frac{J}{\sigma }\right) \sqrt{\frac{\tau -1}{\left(
5-3\tau \right) \pi }}\frac{\Gamma \left[ \frac{1}{\tau -1}\right] }{\Gamma %
\left[ \frac{3-\tau }{2\left( \tau -1\right) }\right] },  \label{29} \\
b_{s} &=&-\frac{4}{3}\left( \frac{J}{\sigma }\right) ^{3}\sqrt{\frac{\left(
\tau -1\right) ^{3}}{\left( 5-3\tau \right) ^{5}\pi }}\frac{\Gamma \left[
\frac{1}{\tau -1}\right] }{\Gamma \left[ \frac{5-3\tau }{2\left( \tau
-1\right) }\right] }, \\
c_{s} &=&\left( \frac{J}{\sigma }\right) ^{5}\frac{\tau }{5}\sqrt{\frac{\tau
-1}{\left( 5-3\tau \right) ^{5}\pi }}\frac{\Gamma \left[ \frac{1}{\tau -1}%
\right] }{\Gamma \left[ \frac{3-\tau }{2\left( \tau -1\right) }\right] }.
\end{eqnarray}

\begin{figure}[t]
\begin{center}
\vspace{1.0cm} \includegraphics[width=0.75\textwidth,angle=0]{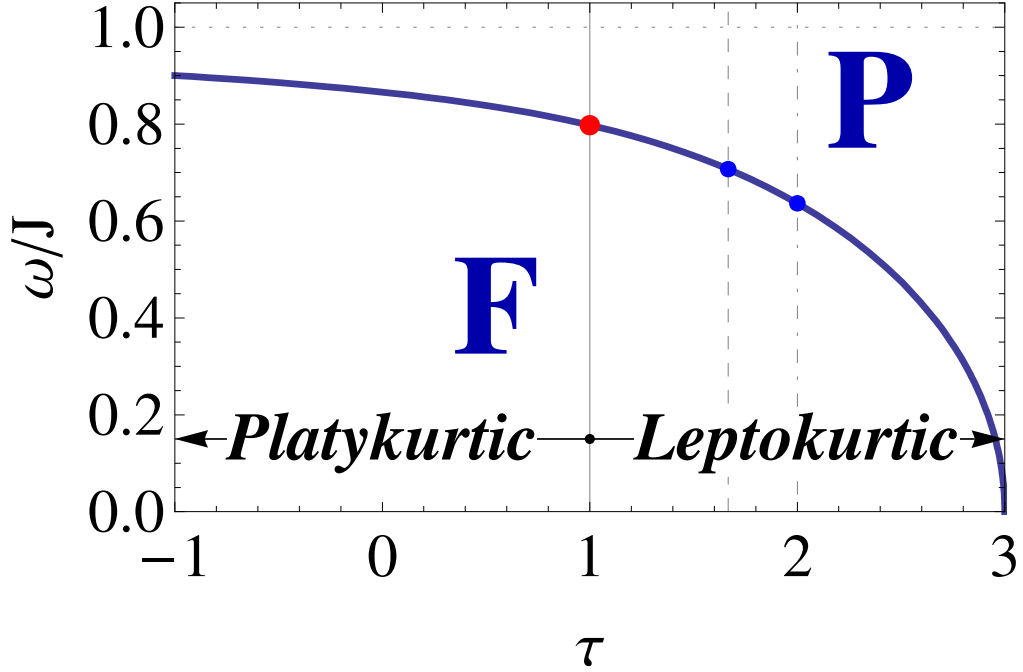}
\end{center}
\caption{(Color online) Zero-temperature phase diagram separating the Ferromagnetic (%
\textbf{F}) and the Paramagnetic (\textbf{P}) phases for platykurtic ($%
\protect\tau <1$) and leptokurtic ($\protect\tau >1$) distributions. The horizontal dotted line represents the limiting case $\protect\tau \rightarrow -\infty $,
\textit{i.e.}, the uniform distribution ($\protect\omega /J=1$), the dashed
vertical line represents the limit for finite standard deviation ($\protect%
\tau =5/3$) and the dot-dashed line the limit for finite average ($\protect%
\tau =2$). The points emphasize the intersection between vertical lines and
the critical line. They correspond to values of $\protect\omega /J$ equal to
$\protect\sqrt{2/\protect\pi}$, $\protect 1/\sqrt{2}$ and $2/%
\protect\pi $, respectively.}
\label{Fig4}
\end{figure}

\noindent The continuous critical frontier at zero temperature is obtained
for $a_{s}=1 $,
\begin{equation}
\frac{\sigma }{J}=\frac{2}{\sqrt{\pi }}\sqrt{\frac{\tau -1}{\left( 5-3\tau
\right)}}\frac{\Gamma \left[ \frac{1}{\tau -1}\right] }{\Gamma \left[ \frac{%
3-\tau }{2\left( \tau -1\right) }\right] }~,  \label{30}
\end{equation}%
\noindent as long as $b_{s}<0$. In the range $1<\tau <5/3$, we notice that
the coefficient $b_{s}$ is always negative, indicating the occurrence of
continuous phase transitions for all values of $\tau $. This expression
permit us to determine the values of $\sigma/J$, or equivalently, the values of $\omega/J$ [see Eq. (\ref{5})] at $T=0$ of the phase diagrams depicted in Fig. \ref{Fig3}. In Fig. \ref{Fig4}, we show the zero-temperature phase diagram, on the plane $\omega /J$ vs $\tau $.

%%%%%%%%%%%%%%%%%%%%%%%%%%%%%%%%%%%%%%%%%%%%%%%%%%%%%%%%%%%%%%%%%%%%%%%%
%%%%%%%%%%%%%%%%%%%%%%%%%%%%%%%%%%%%%%%%%%%%%%%%%%%%%%%%%%%%%%%%%%%%%%%%%

\subsubsection{Finite width: $5/3<\protect\tau <3$}

Mark that in this range we must use $\omega $. Thus, as previously, the free
energy and magnetization are respectively,%
\begin{eqnarray}
f_{s} &=&\sqrt{\frac{\tau -1}{\left( 3-\tau \right) \pi }}\frac{\Gamma \left[
\frac{1}{\tau -1}\right] }{\Gamma \left[ \frac{3-\tau }{2\left( \tau
-1\right) }\right] }\,\omega \,\left\{ \frac{(3-\tau )}{(2-\tau )}\left( 1-%
\frac{1-\tau }{3-\tau }\left( \frac{J}{\omega }\right) ^{2}m_{s}^{2}\right)
^{\frac{1}{1-\tau }+1}+\right.  \nonumber  \label{fefinite} \\
&&\left. 2\left( \frac{J}{\omega }\right) ^{2}\,_{2}F_{1}\left( \frac{1}{2},%
\frac{1}{\tau -1};\frac{3}{2};-\frac{1-\tau }{3-\tau }\left( \frac{J}{\omega
}\right) ^{2}m_{s}^{2}\right) m_{s}^{2}\right\} ,
\end{eqnarray}%
and
\begin{equation}
m_{s}=2\,m_{s}\left( \frac{J}{\omega }\right) \sqrt{\frac{\tau -1}{\left(
3-\tau \right) \pi }}\frac{\Gamma \left[ \frac{1}{\tau -1}\right] }{\Gamma %
\left[ \frac{3-\tau }{2\left( \tau -1\right) }\right] }\;_{2}F_{1}\left[
\frac{1}{2},\frac{1}{\tau -1};\frac{3}{2};\frac{\left( \tau -1\right) J^{2}}{%
\left( \tau -3\right) \omega ^{2}}m_{s}^{2}\right] .  \label{31}
\end{equation}%
Analogously to the above cases, we expand the magnetization $m_{s}$, Eq.~(%
\ref{31}), in powers of $m_{s}$,
\begin{equation}
m_{s}=a_{s}m_{s}+b_{s}m_{s}^{3}+c_{s}m_{s}^{5}+\mathcal{O}(m_{s}^{7}),
\label{32}
\end{equation}%
with the coefficients,
\begin{eqnarray}
a_{s} &=&2\left( \frac{J}{\omega }\right) \sqrt{\frac{\tau -1}{\left( 3-\tau
\right) \pi }}\frac{\Gamma \left[ \frac{1}{\tau -1}\right] }{\Gamma \left[
\frac{3-\tau }{2\left( \tau -1\right) }\right] },  \label{35} \\
b_{s} &=&-\frac{2}{3}\left( \frac{J}{\omega }\right) ^{3}\sqrt{\frac{1}{\pi %
\left[ (4-\tau )\tau -3\right] }}\frac{\Gamma \left[ \frac{1}{\tau -1}\right]
}{\Gamma \left[ \frac{3-\tau }{2\left( \tau -1\right) }\right] }, \\
c_{s} &=&\frac{2}{45}\left( \frac{J}{\omega }\right) ^{5}\sqrt{\frac{\tau -1%
}{\left( 3-\tau \right) \pi }}\frac{(5+9\tau )}{(3-\tau )^{2}}\frac{\Gamma %
\left[ \frac{1}{\tau -1}\right] }{\Gamma \left[ \frac{3-\tau }{2\left( \tau
-1\right) }\right] },
\end{eqnarray}

Thus, the continuous critical frontier is given by
\begin{equation}
\frac{\omega }{J}=\frac{2}{\sqrt{\pi }}\left( \frac{\tau -1}{3-\tau }\right)
^{1/2}\frac{\Gamma \left[ \frac{1}{\tau -1}\right] }{\Gamma \left[ \frac{%
3-\tau }{2(\tau -1)}\right] }~,  \label{36}
\end{equation}%
\noindent and we have again verified that $b_{s}<0$ for all values $5/3<\tau
<3$. We can see in Fig. \ref{Fig4} the zero-temperature phase diagram in the plane containing the width $\omega $ and the generalized degree of freedom $\tau $.

In this case, it is worth noticing an important result. In the free energy
per spin~(\ref{fefinite}), the integrals are finite only for $\tau <2$,
\textit{i.e.}, the free energy at temperature equal to zero is not finite
for probability density functions broader than the Cauchy-Lorentz. Although
we do not have an unequivocal physical account for this phenomenon, we
introduce some insight into this result with the help of the statistical
meaning of our distributions. As mentioned in section 2, for $\tau >1$,
distribution (\ref{3}) is understood as a generalization of Student-$t$ for
real degrees of freedom according to the relation,%
\begin{equation}
\tau =\frac{3+m}{1+m}.
\end{equation}

%%%%%%%%%%%%%%%%%%%%%%%%%%%%%%%%%%%%%%%%%%%%%%%%%%%%%%%%%%%%%%%%%%%%%%%%
\begin{figure}[t]
\begin{center}
\hspace{0.5cm} \includegraphics[width=0.85\textwidth,angle=0]{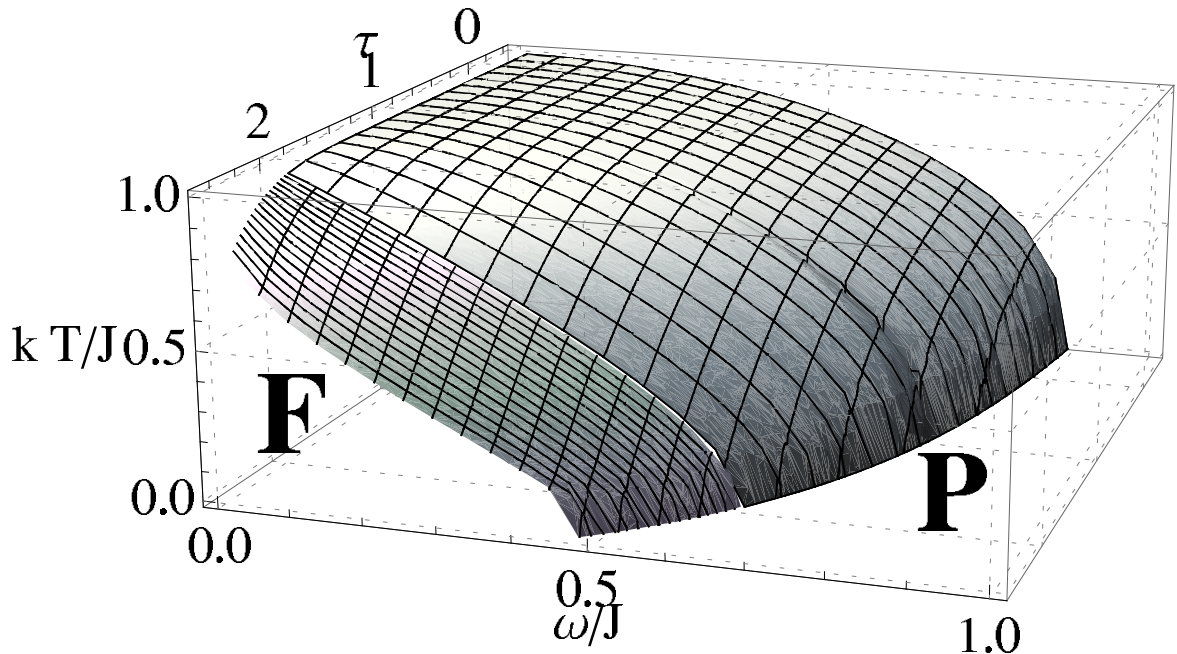}
\end{center}
\caption{(Color online) Tri-dimensional phase diagram of the model in the
axis temperature, $\protect\tau $ and $\protect\omega /J$, separating the
Ferromagnetic (\textbf{F}) and the Paramagnetic (\textbf{P}) phases. We have
used a darker color to represent the regime of $\protect\tau $  ($\tau>2$) in which we
have determined a divergent free energy per spin analytically found at $T=0$%
. }
\label{Fig5}
\end{figure}
%%%%%%%%%%%%%%%%%%%%%%%%%%%%%%%%%%%%%%%%%%%%%%%%%%%%%%%%%%%%%%%%%%%%%%%%%

The Cauchy-Lorentz distribution, Eq. (\ref{3}) with $\tau =2$, corresponds
to the case for which the distribution presents a divergence in the average
but a null average value of the corresponding escort distribution. The
divergence of the mean value of the free energy for $\tau >2$ emerges from
that feature of the property of Eq. (\ref{3}). Moreover, this divergence was
experimentally observed in organic charge-transfer compounds \cite%
{fat-materials}.

In order to summarize the results presented in the manuscript, we show in
Fig. \ref{Fig5} a tridimensional phase diagram separating the ferromagnetic
(\textbf{F}) and the paramagnetic (\textbf{P}) phases defined by the axis
temperature (in units of $J$), $\tau $ and $\omega$ (also in unit of $J$). We observe a contraction of the ferromagnetic phase for increasing values of $\tau $. We have spotted the above-described change in the concavity of the critical frontier for $\tau >2$, as well as the dwindling of the ferromagnetic phase (for increasing values of $\tau $)
which in limit $\tau \rightarrow 3$ turns into the point $\omega =0$.

%%%%%%%%%%%%%%%%%%%%%%%%%%%%%%%%%%%%%%%%%%%%%%%%%%%%%%%%%%%%%%%%%%%%%

\section{Concluding remarks}

In this work we have investigated the infinite-range-interaction Ising model
in the presence of a random magnetic field following a family of continuous
probability density functions, defined by a parameter $\tau $ comprising the
$r$-distribution, for $\tau <1$, and the Student-$t,$ for $\tau >1$, which
have already found their statistical relevance within other contexts of
disordered systems. Moreover, specific PDFs like the Gaussian ($\tau
\rightarrow 1$), the uniform ($\tau \rightarrow -\infty $) and the
Cauchy-Lorentz ($\tau =2$) are obtained thereof. Independently of $\tau $,
we have observed a continuous phase transition with the lessening of the
ferromagnetic phase in the $\frac{k\,T}{J}$ vs $\frac{\omega }{J}$ plane
that corresponds to the region defined by $0\leq \frac{k\,T}{J}\leq 1$ and $%
0\leq \frac{\omega }{J}\leq 1$ in the uniform case and to the point $\frac{%
\omega }{J}=\frac{k\,T}{J}=0$ for $\tau \rightarrow 3$. For $\tau \geq 2$,
we have noted the appearance of an inflexion point for finite $\frac{\omega
}{J}$ and $\frac{k\,T}{J}$ that is also associated with a divergence of the
free energy per spin at null temperature for which we have provided with an
explanation based on the statistical nature of distributions that are fatter
than the Cauchy-Lorentz.

As an extension of this work, a numerical study by means of Monte Carlo
simulations of the model defined by Eqs. (\ref{1}), (\ref{2}) and (\ref{3})
in the case of nearest-neighbors interactions is though to bring a better
understanding of the physical properties of the Ising model in the presence
of random magnetic fields that follow continuous probability distributions
\cite{part2}.

\vskip 2\baselineskip

{\large \textbf{Acknowledgments}}

\vskip \baselineskip
\noindent The authors acknowledge F.~D. Nobre for discussions on several
aspects of disordered magnetic systems. SMDQ benefits from financial support
from the European Union's Marie Curie Fellowship Programme and NC and DOSP
thank the financial support from the Brazilian agency CNPq.

\end{document}